\documentclass[osajnl,twocolumn,showpacs,superscriptaddress,10pt]{revtex4-1} 
\usepackage{amsmath,amssymb,graphicx}
\begin{document}

\title{A Raman Waveplate for Spinor BECs}

\author{Justin T. Schultz}\email{Corresponding author: j.t.schultz@rochester.edu}
\affiliation{The Institute of Optics, University of Rochester, Rochester, NY 14627, USA.}
\affiliation{Center for Coherence and Quantum Optics, University of Rochester, Rochester NY 14627, USA.}

\author{Azure Hansen}
\affiliation{Center for Coherence and Quantum Optics, University of Rochester, Rochester NY 14627, USA.}
\affiliation{Department of Physics and Astronomy, University of Rochester, Rochester, NY 14627, USA.}

\author{Nicholas P. Bigelow}
\affiliation{The Institute of Optics, University of Rochester, Rochester, NY 14627, USA.}
\affiliation{Center for Coherence and Quantum Optics, University of Rochester, Rochester NY 14627, USA.}
\affiliation{Department of Physics and Astronomy, University of Rochester, Rochester, NY 14627, USA.}

\begin{abstract}We demonstrate a waveplate for a pseudo-spin-1/2 Bose-Einstein condensate using a two-photon Raman interaction. The angle of the waveplate is set by the relative phase of the optical fields, and the retardance is controlled by the pulse area. The waveplate allows us to image maps of the Stokes parameters of a Bose-Einstein condensate and thereby measure its relative ground state phase. We demonstrate the waveplate by measuring the Stokes parameters of a coreless vortex. 
\end{abstract}


\maketitle 
Spinor Bose-Einstein condensates (BECs) have proven to be a useful system for creating analogs of phenomena that appear in other systems. The spinor wavefunction can be engineered with the use of external optical and magnetic fields to produce synthetic gauge fields \cite{Gauge1,Gauge2}, vortices \cite{Kevin1}, skyrmions \cite{LSLeslie2009,Skyrmion,Skyrmion2}, and Dirac monopoles \cite{monopoles}. Along with demonstrating the creation of these analogs, new experimental techniques are needed to characterize these systems and study the evolution of not only the populations of the spin ground states but especially the relative phases of these states. 

One possible approach for achieving these goals is through the field of atom optics. Many atom-optical systems have already been realized such as atomic waveguides, lenses, beamsplitters, and atom interferometers (see \cite{Meystre, Cronin} and references therein). Using BECs as the atomic medium allowed for the exploration of coherent atom optics with atom lasers and nonlinear processes such as four-wave mixing. Extending atom optics to spinor BECs allows us to explore the coherent atom-optics analog of optical polarization and provides an avenue for finding the relative ground state phase for the particular case of a pseudo-spin-1/2 system.

\begin{figure}[htbp]
\centerline{\includegraphics{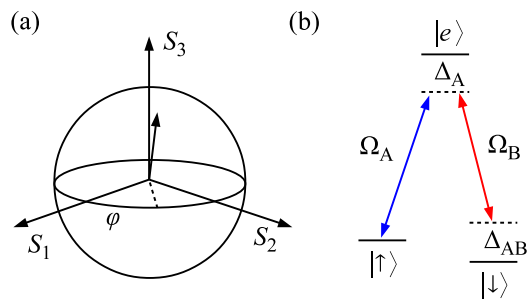}}
\caption{(a) The Bloch Sphere. (b) Three-level Lambda system.}
\label{LambdaSystem}
\end{figure}

In this Letter, we demonstrate a waveplate for pseudo-spin-1/2 BEC and use it to measure the ground state phase via the atomic analog of the optical Stokes parameters. In 1936 Beth \cite{Beth} and Holbourn \cite{Holbourn} showed that light with circular polarization carries spin angular momentum, which leads to the realization that circular polarization states are also spin eigenstates \cite{Fano-Stokes}. The energy eigenstates of a two-level atom and transverse polarization of light are both spin-1/2 Bose systems. We make the correspondence that the amplitude of a given atomic spin eigenstate ($a_{i},i\in\{\uparrow,\downarrow\}$) is analogous to the amplitude of one spin eigenstate of the optical electric field. Any general normalized state in the system can be represented geometrically as a point $(S_1,S_2,S_3)$ on the surface of a sphere of unit radius (Fig. \ref{LambdaSystem}(a)). 
When describing polarization, the sphere is called the Poincar\'{e} sphere \cite{Jerrard}, and for an atomic system, it is termed the Bloch sphere \cite{Allen}. The points on the sphere representing the spin eigenstates (right-hand and left-hand circular polarization) are antipodal and correspond to the points on the Bloch sphere which represent the spin eigenstates (traditionally the North and South poles). The coordinates $S_k$ are called the Stokes parameters and completely describe the spin state up to a global phase. The Stokes parameters of an arbitrary atomic state $|\psi\rangle  = c_{\uparrow}|\uparrow\rangle +e^{i\varphi}c_{\downarrow}|\downarrow\rangle$ (with $0\leq c_{\uparrow , \downarrow}\in\mathbb{R}$) are expressed as \begin{eqnarray}
S_0 &=&c_{\uparrow}^2 + c_{\downarrow}^2\\
S_1 &=&c_{\uparrow}c_{\downarrow}\cos\varphi \\
S_2 &=&c_{\uparrow}c_{\downarrow}\sin\varphi\\
S_3 &=&c_{\uparrow}^2 - c_{\downarrow}^2. 
\end{eqnarray} For an unnormalized state, $S_0\neq 1$, and the other parameters can be normalized by dividing each one by $S_0$. It is straightforward to see that Stokes parameters reveal not only the spin state amplitudes but also the relative phase, $\varphi$.  

Here, we demonstrate that a two-photon Raman interaction can be used to extract the Stokes parameters for a pseudo-spin-1/2 BEC. Consider the three-level Lambda system shown in Fig. \ref{LambdaSystem}(b). Two simultaneous plane waves with Rabi frequencies $\Omega_A$ and $\Omega_B$ are incident on the atoms. In the rotating frame, the dynamics are described \cite{Kevin3, Shore} by 
\begin{eqnarray}
\dot{a}_{\uparrow} &=& -\frac{i}{2}\Omega^*_{A}a_e\\
\dot{a}_{\downarrow} &=& -\frac{i}{2}\Omega^*_B a_e +i\Delta_{AB}a_{\downarrow}\\
\dot{a}_e &=& -i\Delta_A a_e-\frac{i}{2}\Omega_{A}a_{\uparrow}-\frac{i}{2}\Omega_{B} a_{\downarrow}.
\end{eqnarray}
The state amplitudes are given by $a_i$ with $c_i = |a_i|$ and $\varphi = \arg (a_{\downarrow}/a_{\uparrow})$. The two- and one-photon detunings are $\Delta_{AB}$ and $\Delta_A$, respectively. The Raman process requires that the two-photon detuning be zero ($\Delta_{AB} = 0$). For a large detuning $\Delta_A$, population does not enter the excited state, so we may adiabatically eliminate the excited state thereby making the three-level atom a pseudo-spin-1/2 system \cite{AdiabaticElim1,AdiabaticElim2}. For square optical pulses, the Rabi frequencies are constant over the interaction time and the equations can be integrated directly. With equal Rabi frequencies ($|\Omega_{A}|=|\Omega_{B}|$), the Raman interaction can be described via
\begin{equation} \label{RamanWaveplateEq}
\vec{\psi}(t) =e^{i\frac{\Omega t}{2}}\left[\begin{array}{cc} \cos\frac{\Omega t}{2} & ie^{-i\phi}\sin\frac{\Omega t}{2}\\ie^{i\phi}\sin\frac{\Omega t}{2} & \cos\frac{\Omega t}{2} \end{array} \right]\vec{\psi}(t=0)
\end{equation}
with $\Omega = (|\Omega_A|^2+|\Omega_B|^2)/4\Delta_A$. The matrix describing the Raman interaction is identical to the Jones matrix for an arbitrary optical waveplate in a circular basis \cite{Jerrard}. The retardance is given by the pulse area $\Omega t$, and the waveplate angle, $\phi/2=(\phi_A-\phi_B)/2$, is given by half the relative phase between the Raman beams with Rabi frequencies $\Omega_j = |\Omega_j|e^{i\phi_j}$. 

A first example of this analogy is the familiar Rabi flopping of a two-level system illuminated by a cw field \cite{Allen}.  For half a Rabi cycle, if population began completely in $|\uparrow\rangle\,(|\downarrow\rangle)$ it transfered completely to $|\downarrow\rangle\,(|\uparrow\rangle)$ and  $\Omega t = \pi$. In this case, the retardance of the Raman waveplate is $\pi$, and the interaction is analogous to a half-wave plate. 

To demonstrate our ability to measure the relative phase of the ground states, we use this technique to create 2D maps of the Stokes parameters of a coreless vortex. In cylindrical coordinates $(\rho, \varphi)$, the wavefunction of an ideal coreless vortex is of the form $|\psi(\rho,\varphi)\rangle = |c\rangle+|v\rangle =\sqrt{n(\rho)}(|\uparrow\rangle+\rho^{|\ell|}\exp(i\ell\varphi)|\downarrow\rangle)$ where $n(\rho)$ is the column density of the cloud and $\ell$ is an integer describing the vortex winding number. The azimuthally varying phase of the $|v\rangle  =\sqrt{n(\rho)}\rho^{|\ell|}\exp(i\ell\varphi)|\downarrow\rangle$ state shows it is a vortex, and the core of the vortex is filled with population in the non-vortex $|c\rangle = \sqrt{n(\rho)}|\uparrow\rangle$ state. The relative phase of the two states is simply the azimuthally varying phase $\ell\varphi$ which changes from 0 to $2\pi\ell$. 

Our method to create a coreless vortex is described in detail in \cite{Kevin1}. Briefly, a BEC of $6.5\times 10^{6}$ $^{87}$Rb atoms is created in the $|\uparrow\rangle = |F=2,m_f=2\rangle$ state. Gaussian and Laguerre-Gaussian optical modes with orthogonal circular polarization detuned $440\,$MHz from the $|e\rangle = |F'=1,m_f'=1\rangle$ state couple $|\uparrow\rangle$ and $|\downarrow\rangle = |F=2,m_f=0\rangle$ via the $^{87}$Rb D$_1$ transition. Atoms that change state from $|\uparrow\rangle$ to $|\downarrow\rangle$ acquire the azimuthal phase difference between the optical modes, thereby creating a coreless vortex. 

\begin{figure}[h!]
\centerline{\includegraphics{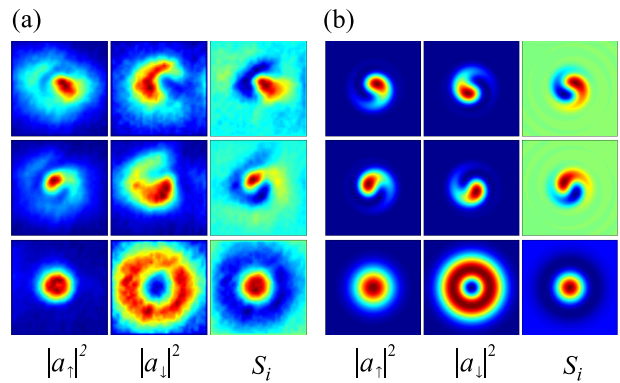}}
\caption{A polarizing beamsplitter for atoms. The first two columns of (a) are averages of 2-3 highly similar absorption images of the outputs of the atom polarizing beamsplitter ($|a_{\uparrow}|^2\,,|a_{\downarrow}|^2$) for essentially identical coreless vortices. The first two rows correspond to $\phi = \pi/2$ and $0$ of the Raman waveplate, and the third row is without the waveplate operation. The final column shows the Stokes maps $S_i$\, $i\in\{1,2,3\}$, which are the differences of the two beamsplitter outputs. The $S_i$ are arranged such that the index $i$ corresponds to the row number. (b) shows the corresponding theory when the coreless vortex is created with a radial phase.}
\label{BSplitter}
\end{figure}

To implement the Raman waveplate, two simultaneous $5\,\mu$s square pulses detuned $\Delta_A= 440\,$MHz below resonance with Gaussian spatial modes couple the $|\uparrow\rangle$ and $|\downarrow\rangle$ states through $|e\rangle$. These Gaussian beams with orthogonal circular polarizations are  copropagating along the $11$ Gauss quantization axis and are derived from the same laser. The pulses are sculpted with acousto-optic modulators which control both $\Delta_A$ and $\Delta_{AB}$. The Gaussian spatial profiles are large with respect to the size of the atomic cloud at the time of the interaction, making them approximately plane waves. The relative phase of the optical beams is adjustable via a controllable optical delay in one beam path after which they share a common optical path. The phase acquired in the optical delay path is locked interferometrically to an accuracy of $\pi/50$. The optical power is optimized by transferring half the population of a spin-polarized BEC in the $|\uparrow\rangle$ state to the $|\downarrow\rangle$ state. In this configuration $\Omega t = m\pi/2$ for an odd integer $m$, and Eq. \ref{RamanWaveplateEq} is that of a quarter waveplate. Shot-to-shot intensity fluctuations on the order of 5\% limit the accuracy of retartance, $\Omega t$, to within $\pi/40$. Via the Stern-Gerlach effect, a magnetic field gradient spatially separates the atomic spin states which are then imaged with resonant absorption imaging. 

Without the Raman waveplate interaction, the magnetic field gradient acts as an equivalent polarizing beamsplitter that separates the spin eigenstates. Applying this Stern-Gerlach process to a coreless vortex separates the core $|c\rangle$ from the vortex $|v\rangle$, giving both $S_0$ and $S_3$, which are the sum and difference of images of the two density distributions, respectively. The waveplate interaction changes the effective measurement basis of the magnetic field gradient. The vortex and core states are superimposed with a relative phase determined by the relative phase of the Raman beams, $\phi$. For $\phi=0\, (\pi/2)$ the Stern-Gerlach beamsplitter measures in the equivalent of the diagonal/anti-diagonal (horizontal/vertical) basis, which enables the measurement of $S_2$ ($S_1$) by the subtraction of images of the two density distributions.

The results of these measurements appear in Fig. \ref{BSplitter}(a). Since the imaging is destructive, a single coreless vortex produces a single Stokes parameter in addition to $S_0$. We can reliably create nearly identical coreless vortices to acquire all three Stokes parameters that very accurately represent the Stokes parameters of any one of the coreless vortices prepared. As the relative phase of the Raman beams is varied, the atomic interference pattern rotates because the magnetic field gradient measures in a basis dependent upon $\phi$. The Stokes parameters $S_1$ and $S_2$ do not look as simple as one would expect from the ideal form of a coreless vortex given above. The radial intensity gradient of the Laguerre-Gaussian beam adds a radially varying phase due to the AC Stark shift as explained in \cite{Kevin2}. In Fig. \ref{BSplitter}(b) the radial phase has been added to the model to show good agreement with the measured Stokes parameters. 
\begin{figure}[h!]
\centerline{\includegraphics{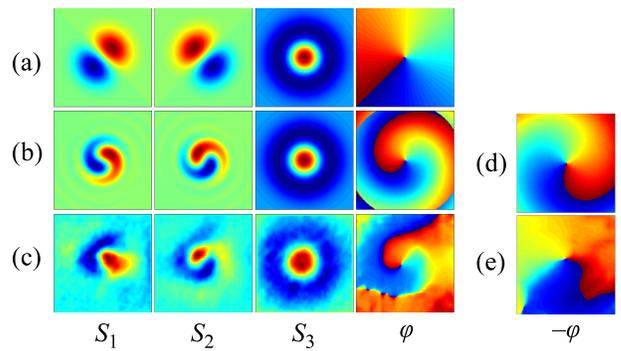}}
\caption{Stokes maps and relative ground state phase of (a) an ideal coreless vortex (theory), (b) a coreless vortex with a radial phase (theory), and (c) an experimentally created coreless vortex. (d) shows a theoretical relative ground state phase for a coreless vortex with opposite handedness to that in (b), and (e) is the relative ground state phase of the experimentally created coreless vortex.}
\label{StokesMaps}
\end{figure}

Reversing the handedness of the Laguerre-Gaussian beam also reverses the rotation direction of the atomic interference pattern and therefore the Stokes parameters. The relative ground state phase, $\varphi$, of the coreless vortex can be found using $S_1$ and $S_2$ via $\varphi = \arctan(S_2/S_1)$. The phases of the coreless vortices with $\ell = \pm 1$ show the radial component arising from the Laguerre-Gaussian intensity pattern during the coreless vortex creation process (see Fig. \ref{StokesMaps}). 

In conclusion, we have extended atom optics to study analogs of optical polarization. Atomic spin states in a spinor BEC are the analogs of polarization states of light, and a two-photon Raman interaction can be used as a waveplate to measure the equivalent Stokes parameters and therefore the relative ground state phase of the spin states. We have demonstrated a Raman waveplate by measuring 2D maps of the Stokes parameters and the relative ground state phase of a coreless vortex. 

This technique is a useful tool to study atom-optic analogs of optical beams with spatially varying transverse polarizations \cite{FPBeam, Milione} and to investigate the evolution of matter-wave phases \cite{Gouy1,Gouy2}, particularly the evolution of the relative ground state phase when the BEC is prepared in regimes with non-trivial spin interactions. Because atoms can have higher dimensional spin manifolds, there is the potential to study richer phenomena that cannot be created in the transverse polarization of optical beams. Although we have demonstrated this Raman waveplate on a BEC, it can, in principle, be used on any three-level system that can be addressed with optical fields. The Raman waveplate allows any arbitrary superposition of two spin states, thereby having possible applications in quantum computing as quantum gates for a single qubit. 

We thank Prof. Miguel Alonso for stimulating discussions. JTS is thankful for an NSF Graduate Research Fellowship. We gratefully acknowledge support from NSF, NASA, and DARPA.


\begin{thebibliography}{99}


\bibitem{Gauge1}J. Dalibard, F. Gerbier, G. Juzeli\={u}nas, and P. \"{O}hberg, ``Colloquium: Artificial gauge potentials for neutral atoms," Rev. Mod. Phys. {\bf 83}, 1523 (2011).  
\bibitem{Gauge2} Y.-J. Lin, R. L. Compton, K. Jim\'{e}nez-García, J. V. Porto, and I. B. Spielman, ``Synthetic magnetic fields for ultracold neutral atoms," Nature {\bf 462}, 628-632 (2009).
\bibitem{Kevin1}K. C. Wright, L. S. Leslie, and N. P. Bigelow, ``Optical control of the internal and external angular
momentum of a Bose-Einstein condensate," Phys. Rev. A {\bf 77}, 041601 (2008).
\bibitem{LSLeslie2009} L. S. Leslie, A. Hansen, K. C. Wright, B. M. Deutsch, and N. P. Bigelow, ``Creation and Detection of Skyrmions in a Bose-Einstein Condensate," Phy. Rev. Lett. {\bf 103}, 250401 (2009).
\bibitem{Skyrmion}J. Choi, W. J. Kwon, and Y. Shin, ``Observation of Topologically Stable 2D Skyrmions in an Antiferromagnetic Spinor
Bose-Einstein Condensate," Phys. Rev. Lett. {\bf 108}, 035301 (2012).
\bibitem{Skyrmion2} J. Choi, S. Kang, S. W. Seo, W. J. Kwon, and Y. Shin, ``Obeservation of a Geometric Hall Effect in a Spinor Bose-Einstein Condensate with a Skyrmion Spin Texture," Phys. Rev. Lett. {\bf 111}, 245301 (2013). 
\bibitem{monopoles} M. W. Ray,	 E. Ruokokoski,	 S. Kandel,	 M. M\"{o}tt\"{o}nen, and D. S. Hall, ``Observation of Dirac monopoles in a synthetic magnetic field," Nature {\bf 505}, 657–660 (2014). 
\bibitem{Meystre}P. Meystre, {\it Atom Optics} (Springer, 2001).
\bibitem{Cronin} A. D. Cronin, J. Schmiedmayer, and D. E. Pritchard, ``Optics and interferometry with atoms and molecules," Rev. Mod. Phys. {\bf 81}, 1051 (2009).
\bibitem{Beth} R. A. Beth, ``Mechanical Detection and Measurement of the Angular Momentum of Light," Phys. Rev. {\bf 50}, 115 (1936). 
\bibitem{Holbourn} A. H. S. Holbourn, ``Angular Momentum of Circularly Polarized Light," Nature {\bf 137}, 31 (1936).
\bibitem{Fano-Stokes} U. Fano, ``A Stokes-Parameter Technique for the Treatment of Polarization in Quantum Mechanics," Phys. Rev. {\bf 93}, 121 (1954). 
\bibitem{Jerrard} H. G. Jerrard, ``Modern description of polarized light: matrix methods," Opt. Laser Technol. {\bf 14}, 309 (1982).
\bibitem{Allen} L. Allen and J. H. Eberly {\it Optical Resonance and Two-Level Atoms} (Dover, 1987).
\bibitem{Kevin3}K. C. Wright, L. S. Leslie, and N. P. Bigelow, ``Raman coupling of Zeeman sublevels in an alkali-metal
Bose-Einstein condensate," Phys. Rev. A {\bf 78}, 053412 (2008).
\bibitem{Shore} B. W. Shore, {\it The Theory of Coherent Atomic Excitations} (Wiley-Interscience, New York, 1990), Volume 2. 
\bibitem{AdiabaticElim1}E. Brion, L. H. Pedersen, and K. M{\o}lmer, ``Adiabatic elimination in a lambda system," J. Phys. A: Math Theor. {\bf 40}, 1033 (2007).
\bibitem{AdiabaticElim2}M. P. Fewell, ``Adiabatic elimination, the rotating-wave approximation and two-photon transitions," Opt. Commun. {\bf 253}, 125 (2005). 
\bibitem{Kevin2}K. C. Wright, L. S. Leslie, A. Hansen, and N. P. Bigelow, ``Sculpting the Vortex State of a Spinor BEC," Phys. Rev. Lett. {\bf 102}, 030405 (2009).
\bibitem{FPBeam} A. M. Beckley, T. G. Brown, and M. A. Alonso, ``Full Poincar\'{e} Beams," Opt. Express {\bf 18}, 10777 (2010).
\bibitem{Milione} G. Milione, H. I. Sztul, D. A. Nolan, and R. R. Alfano, ``Higher-Order Poincar\'{e} Sphere, Stokes Parameters, and the Angular Momentum of Light," Phys. Rev. Lett. {\bf 107}, 053601 (2011). 
\bibitem{Gouy1} I. G. Da Paz, P. L. Saldanha, M. C. Nemes, and J. G. Peixoto de Faria, ``Experimental proposal for measuring the Gouy phase of matter waves," New J. Phys. {\bf 13}, 125005 (2011).
\bibitem{Gouy2} A. Hansen, J. T. Schultz, and N. P. Bigelow, ``Measuring the Gouy Phase of Matter Waves Using Full Bloch Bose-Einstein Condensates," in {\it The Rochester Conferences on Coherence and Quantum Optics and the Quantum Information and Measurement meeting}, OSA Technical Digest (online) (Optical Society of America, 2013), paper M6.64.   



\end{thebibliography}
\end{document}